\documentclass[./sn-mathphys-num]{sn-jnl}

\usepackage{graphicx}%
\usepackage{multirow}%
\usepackage{amsmath,amssymb,amsfonts}%
\usepackage{amsthm}%
\usepackage{mathrsfs}%
\usepackage[title]{appendix}%
\usepackage{xcolor}%
\usepackage{textcomp}%
\usepackage{manyfoot}%
\usepackage{booktabs}%
\usepackage{algorithm}%
\usepackage{algorithmicx}%
\usepackage{algpseudocode}%
\usepackage{listings}%
\usepackage{natbib}
\usepackage{wrapfig}
\usepackage{subcaption}
\usepackage{caption}
\usepackage{float}

\begin{document}

\title[Ultrafast Deep Learning-Based Scatter Estimation in Cone-Beam Computed Tomography]{Ultrafast Deep Learning-Based Scatter Estimation in Cone-Beam Computed Tomography}


\author*[1,2]{\fnm{Harshit} \sur{Agrawal}}\email{harshit.agrawal@aalto.fi}

\author[1]{\fnm{Ari} \sur{Hietanen}}

\author[2]{\fnm{Simo} \sur{S\"{a}rkk\"{a}}}

\affil*[1]{\orgdiv{Department of Electrical and Automation Engineering}, \orgname{Aalto University}, \orgaddress{\postcode{02150}, \city{Espoo}, \country{Finland}}}

\affil[2]{\orgdiv{Research \& Technology}, \orgname{Planmeca Oy}, \orgaddress{\street{Asentajankatu 6}, \postcode{00880}, \city{Helsinki}, \country{Finland}}}


\abstract {\textbf{Purpose:} Scatter artifacts drastically degrade the image quality of cone-beam computed tomography (CBCT) scans.  Although deep learning-based methods show promise in estimating scatter from CBCT measurements, their deployment in mobile CBCT systems or edge devices is still limited due to the large memory footprint of the networks. This study addresses the issue by applying networks at varying resolutions and suggesting an optimal one, based on speed and accuracy.\\
\textbf{Methods:} First, the reconstruction error in down-up sampling of CBCT scatter signal was examined at six resolutions by comparing four interpolation methods. Next, a recent state-of-the-art method was trained across five image resolutions and evaluated for the reductions in floating-point operations (FLOPs), inference times, and GPU memory requirements.\\
\textbf{Results:} Reducing the input size and network parameters achieved a 78-fold reduction in FLOPs compared to the baseline method, while maintaining comparable performance in terms of mean-absolute-percentage-error (MAPE) and mean-square-error (MSE). Specifically, the MAPE decreased to $3.85\%$ compared to $4.42\%$, and the MSE decreased to $1.34 \times 10^{-2}$ compared to $2.01 \times 10^{-2}$. Inference time and GPU memory usage were reduced by factors of 16 and 12, respectively. Further experiments comparing scatter-corrected reconstructions on a large, simulated dataset and real CBCT scans from water and Sedentex CT phantoms clearly demonstrated the robustness of our method.\\
\textbf{Conclusion}: This study highlights the underappreciated role of downsampling in deep learning-based scatter estimation. The substantial reduction in FLOPs and GPU memory requirements achieved by our method enables scatter correction in resource-constrained environments, such as mobile CBCT and edge devices.}




\maketitle

\section{Introduction}\label{sec1}
Cone-beam computed tomography (CBCT) has emerged as a major imaging modality, offering a cost-effective solution with a relatively lower dose compared to multi-detector CT (MDCT) \cite{faccioli2020cost,jacobs2011dental}. A key advantage of CBCT is its portability which can potentially enhance the accessibility of imaging services to a broader population \cite{shuaib2010introduction,faccioli2020cost}. However, CBCT faces its own obstacles, notably scatter, primarily caused by radiating a large 2D flat-panel detector using a cone-shaped X-ray beam \cite{siewerdsen2001cone}. Scatter adversely affects the quality of CBCT images \cite{joseph1982effects,kalender2007flat} by reducing contrast, causing shading, and leading to inaccuracies in CT numbers, thereby limiting the diagnostic applications of CBCT.

Scatter correction methods can be categorized into hardware-based and software-based approaches. Hardware-based methods include the use of air-gaps, anti-scatter grids, beam-blockers, bow-tie filters, and limiting the field-of-measurement (FOM) \cite{sore1985,kyriakou2007efficiency,zhu09,bootsma2011effects,chen2008feasibility}. However, these methods are not universally accepted as they present their own limitations \cite{ruhrnschopf2011general}. The second category of approaches, software-based, include gold-standard Monte Carlo (MC) simulations to simulate the scatter signal and subtract it from the measured projections. Despite their accuracy, MC-based methods are time-intensive, often taking several minutes, and require an initial accurate reconstruction \cite{badal2009accelerating,bertram2008monte}.

As an alternative to MC simulation-based correction, several deep learning-based U-Net \cite{Unet15} architectures have been proposed for scatter estimation from CBCT projections \cite{maier2019,nomura2019,pinto2023deep, cruz2024task, agrawal2024}. These methods learn the low-frequency scatter from scatter-corrupted projections with training data pairs generated using MC simulations. For instance, Maier et al. \cite{maier2019} introduced DSE-Net for real-time scatter estimation for different tube voltages, anatomies, and noise levels in medical CT. Similarly, Erath et al. \cite{erath2021deep} applied DSE-Net for cross scatter correction in dual source CT. To address the variations in FOM size observed in CBCT scans, Agrawal et al.\cite{agrawal20241,agrawal2024} proposed Aux-Net, which incorporates FOM size information within the encoder of a U-Net. While these methods are generally faster than MC simulations for scatter estimation, their time and computational complexity largely depend on the size of the input image and the number of parameters in the network. Mobile CBCT or edge devices may not have sufficient GPU memory to handle such computations for large size of input projections. Therefore, a fast neural network is crucial for the clinical application of these networks in resource-constrained environments. 

To reduce computational complexity, downsampling the input has been a common approach in scatter correction methods simultaneously leveraging the low-frequency nature of scatter distribution. For example, Maier et al. \cite{maier2019} downsampled the input to $384 \times 256$, while Agrawal et al. \cite{agrawal2024} downsampled all CBCT projections of various sizes to $320 \times 256$ to train a single model for varying FOM settings. Table~\ref{tab:input_sizes} summarizes different strategies for resizing inputs in the deep learning-based scatter estimation literature. Variations in input sizes are evident, and the optimal downsampling method for scatter estimation remains unclear. To date, no study has investigated a suitable downsampling size and corresponding method for this task.

On the other hand, some studies have shown the benefits of bicubic interpolation in chest X-ray classification and face recognition tasks. For chest X-ray classification, a bicubic interpolation-based downsampling performed best \cite{hirahara2021effect} among different interpolation algorithms, achieving maximum classification accuracy at an image size of $64\times64$ compared to the original size of $1024\times1024$ pixels. Similarly, for the face recognition from electronic identity cards, a bicubic downsampling method performed better compared to the other interpolation methods \cite{hindratno2023impact}.

\begin{table}
    \caption{Original pixel sizes of the CBCT projections and the resized input dimensions for various methods in the deep learning-based CBCT scatter correction literature.}
    \centering
    \begin{tabular}{cccc}
    \hline
        method & original size & input size \\
        \hline
         Maier et al. \cite{maier2019} & not-provided &  $384\times256$\\
         Lee et al. \cite{lee2019deep} & $256\times128$ & $256\times128$\\
         Nomura et al. \cite{nomura2019} & $372\times372$ & $372\times372$ \\
         Lalonde et al. \cite{lalonde2020evaluation} & $1024\times1024$ & $512\times512$\\
         Agrawal et al. \cite{agrawal2024} & various sizes & $320\times256$ \\
         Bastida et al. \cite{cruz2024task} & $512\times256$ & $512\times256$ \\
    \hline
    \end{tabular}
    \label{tab:input_sizes}
\end{table}

Therefore, we investigated different resizing methods and their effect on the scatter signal reproduction. Further, we adopted the Aux-Net proposed in Agrawal et al.\cite{agrawal2024} for scatter estimation in CBCT scanners under varying FOM settings. We trained several different models using different input sizes and compared these models using a large, simulated dataset and real CBCT scans. Our contributions are as follows:

\begin{itemize}
    \item We explore various interpolation methods to assess their accuracy in downsampling and reconstructing the upsampled target scatter signal.
    \item We demonstrate that downsampling the network input to an ultralow resolution enables an ultrafast scatter estimation network without compromising performance. Empirically, we identified an optimal resolution that balances speed and accuracy.
    \item We show that the network parameters can be reduced by a factor of three while achieving comparable results.
\end{itemize}

\section{Methods}
In this section, we detail the simulated training and testing datasets, the phantom scans utilized for testing, the interpolation methods employed, the network architecture, and the processes for training and evaluation.
\subsection{Simulated Dataset}
 We used MC simulations following the methods described in \cite{salvat2006penelope, agrawal2023} for voxelized geometry to simulate X-ray trajectories. Primary and scatter CBCT projections were simulated using reconstructions of CT scans. We utilized the publicly available HNSCC-3DCT-RT dataset \cite{bejarano2018head} from the TCIA repository \cite{clark2013cancer} and CT scans of anthropomorphic phantoms. Nine scans were randomly selected for training and four scans for testing from HNSCC-3DCT-RT dataset. Additionally, two CT scans of anthropomorphic phantoms were included in the training set, and the third CT scan was included in the testing set. To augment the training set, four reconstructions in the training dataset and one in the testing dataset were downsampled by 60\%. In total, 15 scans were used for training and 6 scans for testing.

 We simulated the realistic geometry of a Viso G7 CBCT system (Planmeca Oy., Helsinki, Finland) for jaw protocol. For each reconstruction in the training dataset, we simulated 100 projection views with an angular range of 210 degrees. The number of X-ray simulated photons was set to 2,500 per detector pixel for each projection view. The starting energy of each photon was sampled from a X-ray spectrum of a tube with maximum voltage of 100 kV. The total size of projections (active detector area) varied with the size of FOM and each pixel in the projection corresponded to a detector pixel size of 0.278 mm. Neither a bow-tie filter nor grid was included in the simulations. We simulated 18 different FOM sizes for training and 30 different FOM sizes (not included in the training) for testing (see Tables A1 and A2 in Appendix A).  Each simulation was repeated 10 times using a random starting seed for the X-ray photon direction and energy sampling. During training, we averaged different sets of the repeated simulations to obtain 10 different noise levels for the inputs. The target scatter was always the average of 10 repeated simulations. This resulted in a total of $15 \times 100 \times 10 \times 18$ i.e. 270,000 projections for the training. For testing, we simulated 500 projection views per scan with 25,000 photons per detector pixel, resulting in $6 \times 500 \times 30$ i.e. 90,000 projections.

\subsection{Phantom measurements}
For the CBCT scans in testing, we utilized a plastic jar (18 cm diameter), a thin plastic bottle (6.5 cm diameter), and a SedentexCT image quality (IQ) phantom (16 cm diameter). The jar and bottle phantoms were filled with water and scanned for a FOM of $170\times170$ mm. The SedentexCT IQ phantom was scanned with different contrast resolution inserts made of aluminum (Al), polytetrafluoroethylene (PTFE), Delrin, low-density polyethylene (LDPE), and air rods suspended in polymethyl methacrylate (PMMA). These scans were performed with FOMs of $170\times170$ mm and $130\times30$ mm, respectively. All of these scans were acquired at 100 kV and 110 mAs using the Viso G7 device. The scanning geometry was consistent with that used in the simulations.

\subsection{Interpolation methods}
We examined the reproduction error in the scatter signal for four different interpolation methods, following the implementation in PyTorch library \cite{ansel2024pytorch}. We compared the resizing errors to a base size of $320 \times 256$ as used in \cite{agrawal2024}. For this, we resized each input to sizes $160\times128$, $80\times64$, $40\times32$, $20\times16$, and $10\times8$, corresponding to factors of 2, 4, 8, 16, and 32, respectively. The downsampled scatter signal was then upsampled back to the original size. Finally, the MSE is calculated over all the target scatter signals in the training set. The compared interpolation methods are as follows:

Nearest-neighbor interpolation \cite{rukundo2012nearest} selects the value of the nearest pixel to assign to the new pixel.
Given an original image with pixel coordinates $(x, y)$ and a resized image with pixel coordinates $(x', y')$, the nearest neighbor interpolation can be expressed as:

\begin{equation}
I'(x', y') = I(\text{round}(\frac{x'}{s_h}), \text{round}(\frac{y'}{s_w})),
\end{equation}
where \(I\) is the original image, \(I'\) is the resized image, \(s_h\) and \(s_w\) are the scaling factors for height and width, respectively, and
\(\text{round}(\cdot)\) is the rounding function that selects the nearest integer pixel coordinate.
The scaling factors \(\frac{x'}{s_h}\) and \(\frac{y'}{s_w}\) map the coordinates of the resized image back to the coordinates of the original image. The nearest pixel value from the original image is then assigned to the corresponding pixel in the resized image.

In area interpolation \cite{wong1997area}, the output value at each position $(x', y')$ is computed as the average of the input values within the corresponding area given by:
\begin{equation}
    I'(x', y') = \frac{1}{|\Omega(x', y')|} \sum_{(x, y) \in \Omega(x', y')} I(x, y),
\end{equation}
where \( I(x, y) \) is the intensity of the original image, \( I'(x', y') \) is the intensity of the resized image, \( \Omega(x', y') \) represents the region in the input image that maps to the output pixel \( (x', y') \), and \( |\Omega(x', y')| \) is the area (or number of contributing pixels) in the original image.

Bilinear interpolation \cite{lehmann1999survey} is a method for estimating the intensity of an output pixel by performing successive linear interpolations along two perpendicular axes. It utilizes the four nearest pixel values in the input image to compute a weighted average. Mathematically, bilinear interpolation can be expressed as a first-order polynomial fit:

\begin{equation} I(x, y) = \sum_{i=0}^{1} \sum_{j=0}^{1} w_{ij} x^i y^j,
\end{equation}
where \( (x, y) \) represent continuous coordinates within the unit square defined by the four nearest input pixels, and \( w_{ij} \) are coefficients determined by their intensity values. Given an output pixel at non-integer coordinates \( (x', y') \), its interpolated intensity is obtained by evaluating \( I(x, y) \) at \( (x', y') \), effectively reconstructing a smooth transition between discrete intensity values.

Unlike bilinear interpolation, which considers only the four nearest pixel values, bicubic interpolation \cite{keys1981cubic} uses 16 neighboring pixels (4 points and 12
derivatives) to estimate an interpolated value. The bicubic interpolation can be written as:
\begin{equation}
    {I}(x, y) = \sum_{i=0}^{3} \sum_{j=0}^{3} w_{ij}x^{i} y^{j},
\end{equation}
where \( (x, y) \) represent continuous coordinates within the local interpolation region, and \( w_{ij} \) are coefficients determined by the intensities, first derivatives, and mixed second derivatives of the 4 nearest input pixels \cite{keys1981cubic}. Given an output pixel at non-integer coordinates \( (x', y') \), its interpolated intensity is obtained by evaluating \( I(x, y) \) at \( (x', y') \), ensuring a higher-order smooth approximation compared to bilinear interpolation.
\begin{figure}[htbp]
\centering
\includegraphics[width=0.8\textwidth]{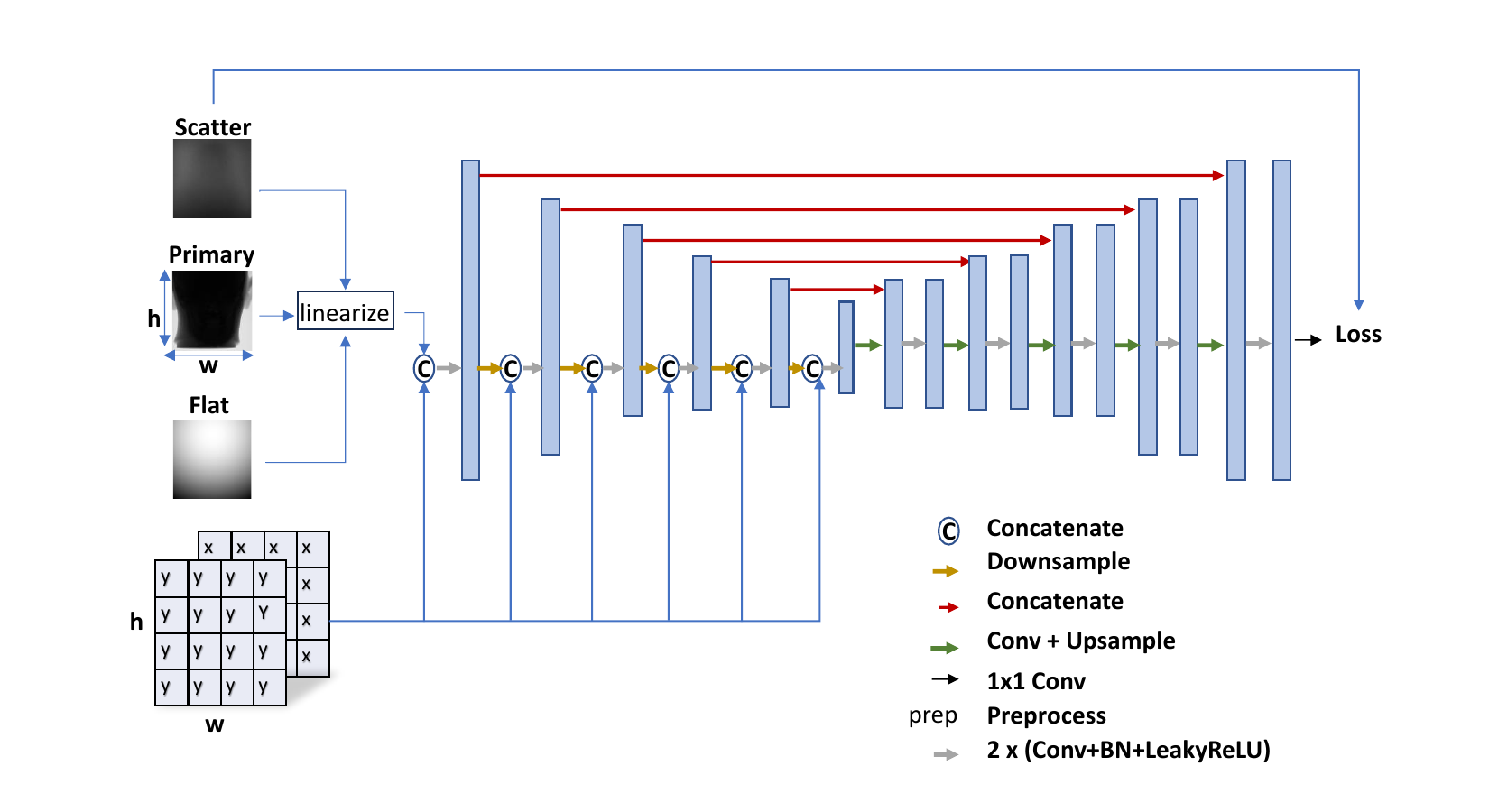}
\caption{The network architecture of Aux-Net \cite{agrawal2024} utilized in the experiments. The additional input channels consists of the normalized width and height, given by $x=\frac{w}{w_{max}}$ and $y=\frac{h}{h_{max}}$.}
\label{net}
\end{figure}
 
\subsection{Network Architecture}
 The Aux-Net architecture is shown in Fig.~\ref{net}, which is a U-Net with additional channels concatenated to the encoder \cite{agrawal2024}. These additional channels contain the normalized height and width of FOM (in mm). There are multiple blocks in the network. The first block has 16 channels, doubling in each of the following blocks. Each block consists of two conv+BN+LeakyReLU layer combination. For net-$320\times256$, $160\times128$, and $80\times64$, the network had five downsampling blocks with a total of 7.3 $\times$ $10^6$
 trainable parameters. While net-$40\times32$ had four downsampling blocks and $20\times16$ had three downsampling blocks with 1.8 $\times$ $10^6$ and 0.5 $\times$ $10^6$ parameters, respectively. The number of the downsampling blocks were reduced to process the reduced input image size. After each block, the features are downsampled by two, using a max pooling layer. The upsampling blocks in the decoder of the network, consists of a convolutional layer followed by a bilinear upsmapling layer.

 \subsection{Training \& Evaluation} We used Adam optimizer with the default parameters with a batch size of 64 to reduce mean-square-error (MSE) loss. A 5-fold cross validation scheme was used for training. Whole training data was divided into 5 folds (20 \% in each fold) and during training, 1-fold (54,000 samples) was used for validation and the remaining 4 folds (216,000 samples) were used for training. The training was run until the validation loss did not decrease for 5 continuous epochs. The initial learning rate was set to $10^{-4}$ and was logarithmically reduced to $10^{-5}$ in 30 epochs and fixed thereafter. All experiments were run on a NVIDIA RTX A4500 card with 20 GB GPU memory using PyTorch library. 
 The baseline model was trained by resizing all the inputs to a size of $320\times256$. Four more models were trained with the resized inputs of sizes $160\times128$, $80\times64$, $40\times32$, and $20\times16$.
 
 During the training, all the input images are resized to a fixed size. During the testing, first, a downsampling layer is inserted in the beginning of the network to downsample the input. Next, the downsampled input is processed by the neural network to estimate the scatter. At the end, the architecture consists of an upsampling layer which upsamples the scatter estimates to the original size of the input. The input to the network was linearized and the target was flat normalized scatter such as:
\begin{equation}
    {P}_{\rm sim} =  -\log \left(\frac{{I}_{\rm sim} + {S}_{\rm sim}}{{F} _{\rm sim}}\right),
\label{eq:3}
\end{equation}
\text{and}  
\begin{equation}
    {S}_{\rm sim} =  -\log \left(\frac{{S}_{\rm sim}}{{F} _{\rm sim}}\right).
\label{eq:6}
\end{equation}
where ${P}_{\rm sim}$ is the linearized projection, ${I}_{\rm sim}$ is the simulated primary, ${S}_{\rm sim}$ is the simulated scatter, and ${F}_{\rm sim}$ is the simulated flat-field projection.

For the simulated test dataset, the corrupt input projections were obtained using Eq.~\eqref{eq:3}, while for the real test scans, the measured projections were linearized using flat-field projections. For the reconstructions, the predicted scatter signal was subtracted from the corrupt input projections and reconstruction was obtained using a filtered back-projection (FDK) algorithm \cite{feldkamp1984practical}. For the simulated data, we calculated MSE and mean-absolute-percentage-error (MAPE) \cite{maier2019} errors for the projection domain scatter predictions, and root-mean-square-error (RMSE) for the scatter-corrected images. Since there is no ground truth for the real scans, we analyzed line profiles from the scatter-corrected axial slices. Additionally, we calculated image uniformity in Hounsfield Units (HU) by determining the absolute difference between the mean of a central region-of-interest (ROI) and the mean of 20 ROIs in the periphery of the water phantom scans.
\section{Results}\label{sec2}
In this section, we first compare the performance of various interpolation methods. Next, we analyze the resource requirements for different trained models. Finally, we present the results of scatter correction on both simulated and real datasets.

\begin{figure}[!ht]
    \centering
    \includegraphics[width=0.99
\linewidth]{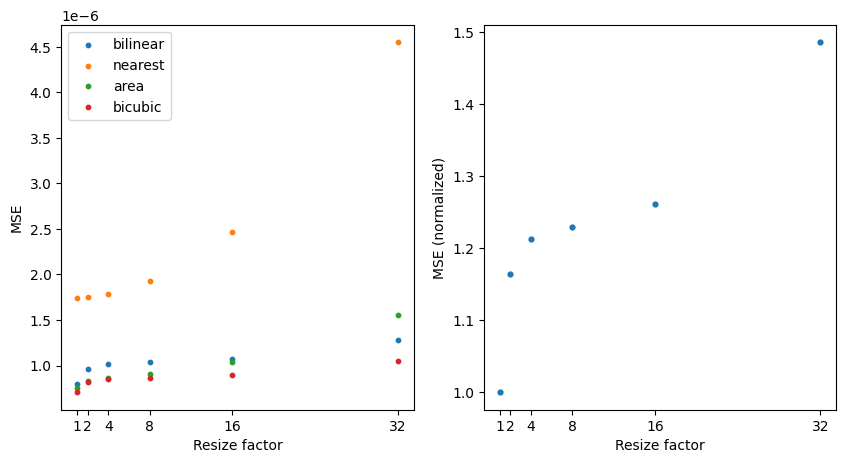}
    \caption{MSE errors calculated for all target scatter images in the training dataset. The resize factors of 1, 2, 4, 8, 16, 32 on the $x-$ axis corresponds to the sizes of $320\times256$, $160\times128$, $80\times64$, $40\times32$, and $20\times16$, respectively. Left: MSE is plotted for four different interpolation methods. Right: MSE for bicubic interpolation is plotted, normalized by the MSE of the size of $320\times256$.}
    \label{fig:interp-error}
\end{figure}
\subsection{Effect of interpolation methods}

The MSE for four interpolation methods, along with the normalized MSE for bicubic interpolation are illustrated in Fig.~\ref{fig:interp-error}. The results indicate that nearest neighbor interpolation yields the highest errors, whereas bicubic interpolation results in the lowest residual error. Furthermore, the scatter reconstruction error increases sharply for a factor of 2 and a factor of 32. Consequently, bicubic interpolation was chosen for downsampling and upsampling the scatter signal in subsequent experiments.

\subsection{Resource Requirements}
The performance of various network configurations evaluated in terms of floating point operations (FLOPs), inference time, memory utilization, and error metrics are summarized in Table \ref{tab:flops}. For an input size of $320 \times 256$, the network configuration net-$320 \times 256$ exhibited the highest FLOPs at 4.71 G and the highest memory utilization during testing at 3890 MB. This configuration achieved a MAPE of $4.42 \pm 0.18\%$ and an MSE of $2.01 \pm 0.14 \times 10^{-2})$. The test time for this configuration was 90 ms.

As the input size decreased, the FLOPs and memory utilization also decreased, considerably. For instance, net-$160 \times 128$ had FLOPs of 1.18 G and test-time memory utilization of 1162 MB, with a MAPE of $3.84 \pm 0.14\%$ and an MSE of $1.56 \pm 0.13 \times 10^{-1}$. The test time for this configuration was 35 ms. Further reductions in input size and model parameters for $80 \times 64$, $40 \times 32$, and $20 \times 16$ continued to show decrease in FLOPs and memory utilization. Notably, net-$40 \times 32$ obtained the lowest MSE of $1.34 \pm 0.09 \times 10^{-2}$ and a comparable MAPE of $3.85 \pm 0.10\%$ under just 5.6 ms.

 \begin{table}
    \centering
     \caption{Resource utilization along with the MAPE and MSE errors. Test time and memory utilization are computed for a dummy input with a batch size of 64 and averaged over 1000 passes through the model. Average MAPE and MSE $\pm$ standard deviation for the projection domain scatter estimation is calculated by averaging the mean MAPE and MSE obtained for each of the 5-folds.}
    \begin{tabular}{ccccccc}
    \hline
        method & params & FLOPs & MAPE & MSE & time & mem\\
        & $(10^6)$ & (G) & (\%) & ($10^{-2}$)& (ms) & (MB)\\
        \hline 
       net-$320\times256$ & 7.3 & 4.71  & 4.42 $\pm$ 0.18 & 2.01 $\pm$ 0.14 & 90 & 3890\\
         net-$160\times128$ & 7.3  & 1.18  & 3.84 $\pm$ 0.14 & 1.56 $\pm$ 0.13& 35& 1162\\
         net-$80\times64$ &  7.3  & 0.29  & 3.85 $\pm$ 0.13 & 1.43 $\pm$ 0.08 & 15& 514\\      
         net-$40\times32$ & 1.8 & 0.06  & 3.85 $\pm$ 0.10 & 1.34 $\pm$ 0.09 & 5.6 & 310\\
         net-$20\times16$ & 0.5 & 0.01  &  4.35 $\pm$ 0.23 & 1.67 $\pm$ 0.21 & 3& 248\\
         net-$20\times16$ & 1.8 &  0.04 &  3.97 $\pm$ 0.15 & 1.50 $\pm$ 0.09 & 5.7 & 272\\
         \hline
    \end{tabular}
   
    \label{tab:flops}
\end{table}

\begin{figure}[!htbp]
\centering
\includegraphics[width=\textwidth]{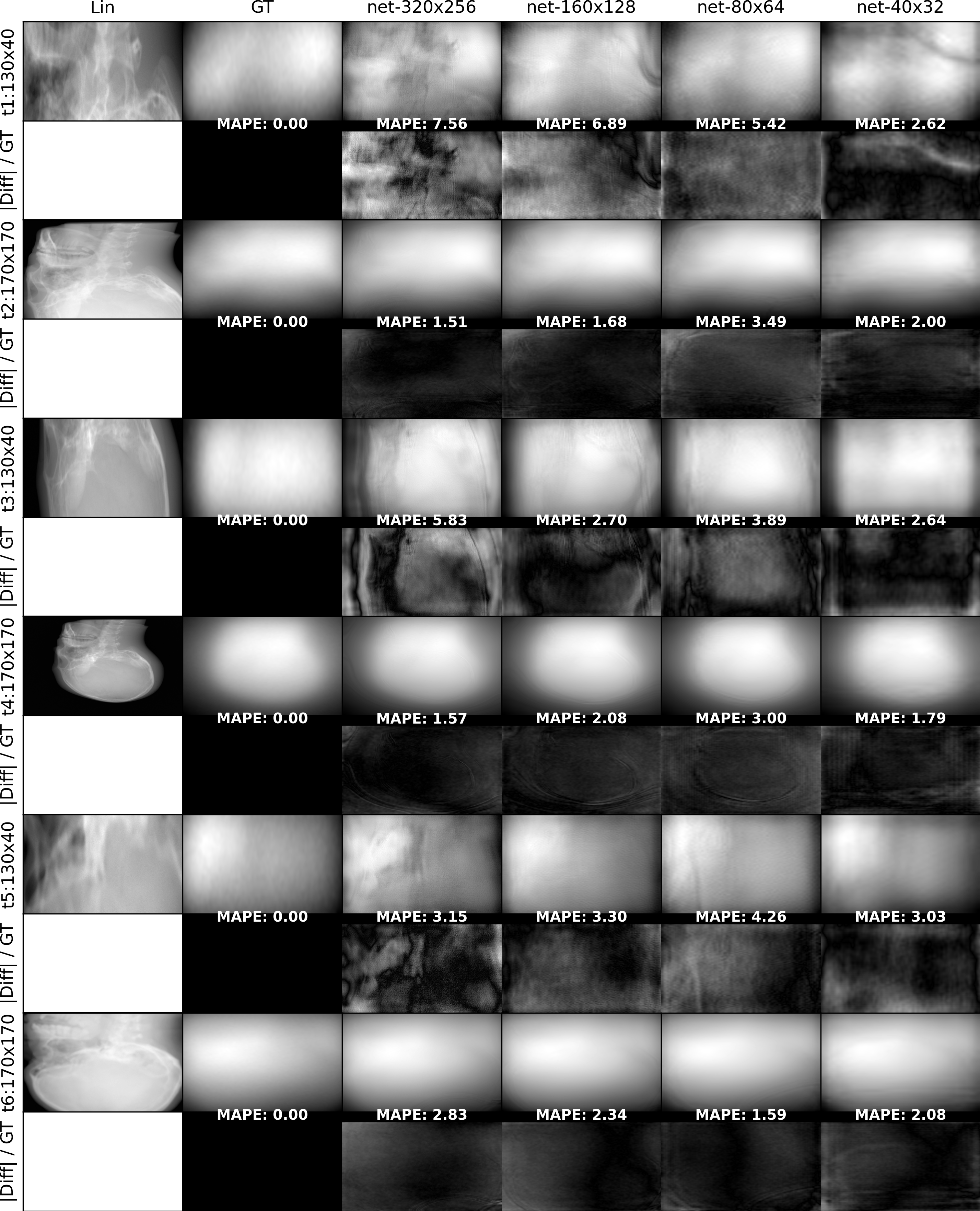}
\caption{Visual and quantitative examples of the estimated scatter, the linearized input, and the ground truth scatter image. The difference images display the relative absolute difference from the ground truth scatter, along with the corresponding MAPE. The display range of the difference images is 0-0.2.}
\label{fig:sim_proj}
\end{figure}

\subsection{Scatter estimation on the simulated projections}
Figure~\ref{fig:sim_proj} compares various models by their MAPE errors in the estimated projection-domain scatter images. Since the MAPE and MSE values did not improve after net-$40\times32$, the images from later networks are not compared. The visualizations include the linearized input, the ground truth scatter, and the absolute error normalized by the ground truth scatter for six test simulations. The net-$40 \times 32$ outperforms the net-$320 \times 160$ for the smallest FOM size of $130 \times 40$. Specifically, the net-$40 \times 32$ achieves MAPEs of 2.62, 2.6, and 3.03, while the net-$320 \times 256$ records MAPEs of 7.56, 5.83, and 3.15, respectively. However, for the largest FOM size of $170 \times 170$ in the t2 and t4 cases, the net-$40 \times 32$ performs slightly worse than the net-$320 \times 256$. 
\begin{figure}[!htbp]
\centering
\includegraphics[width=\textwidth]{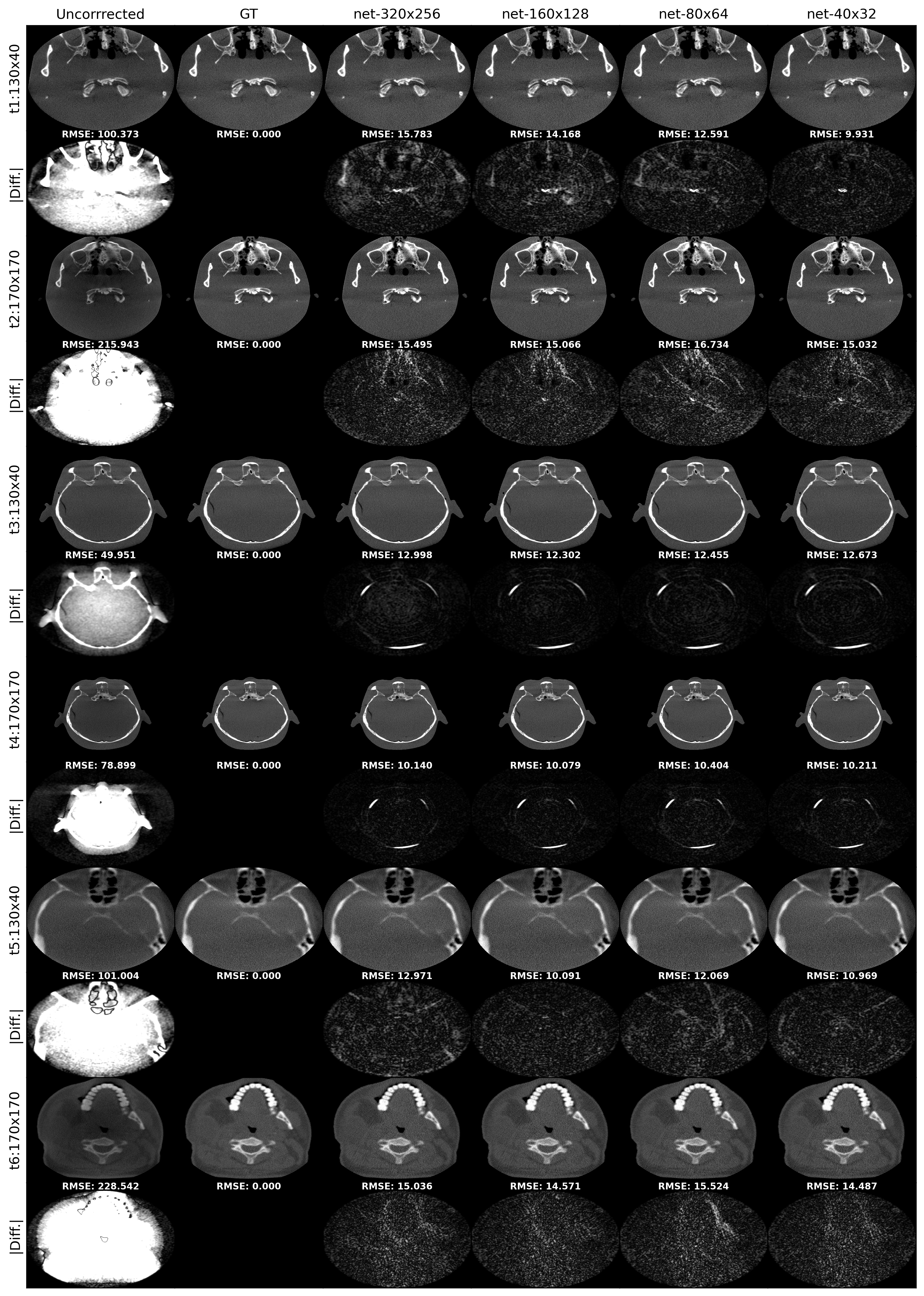}
\caption{Example of the reconstructed axial images after scatter correction along with the uncorrected, ground truth, and the absolute difference from the ground truth images. t:1--6 corresponds to six different test images. The display window center and width of the absolute difference is 50 HU.}
\label{fig:sim_img}
\end{figure}

\begin{table}[t]
\caption{RMSE (HU) for each of the 30 FOM sizes. Last row shows the mean RMSE ($\pm $ standard deviation), calculated over all 30 FOM sizes with the lowest metric emphasized in bold, while the second-lowest metric, underlined.}
\centering   
\begin{tabular}{c|*{5}{c}}
\hline
FOM (mm) & net-$320\times256$ & net-$160\times128$ & net-$80\times64$ & net-$40\times32$ \\
\hline
$130\times40$ & 10.82 & 8.51 & 7.76 & 6.34\\ 
$130\times50$ & 8.90 & 9.10 & 8.34 & 8.20\\ 
$130\times70$ & 8.38 & 8.79 & 9.57 & 9.91\\ 
$130\times80$ & 8.84 & 9.23 & 10.10 & 9.86 \\
$130\times100$ & 9.94 & 10.19 & 10.99 & 10.71\\ 
$130\times110$ & 10.41 & 10.45 & 10.71  & 10.85\\ 
$130\times130$ & 10.85 & 10.41 & 10.79 & 10.74\\ 
$130\times140$ & 11.60 & 10.21 & 10.61 & 11.12 \\
$130\times160$ & 12.03 & 12.49 & 11.16 & 13.05 \\
$130\times170$ & 11.43 & 10.04 & 10.80 & 10.86 \\
\hline

$150\times40$ & 9.80 & 7.43 & 7.06 & 6.06 \\
$150\times50$ & 7.93 & 7.71 & 7.43 & 6.52 \\
$150\times70$ & 7.41 & 7.69 & 7.91 & 8.49 \\
$150\times80$ & 8.15 & 8.23 & 8.32 & 8.64 \\
$150\times100$ & 9.49 & 9.23 & 9.05 & 9.47 \\
$150\times100$ &  9.71 & 9.73 & 9.42 & 9.45 \\
$150\times130$ & 10.82 & 9.70 & 9.64 & 9.75 \\
$150\times140$ & 11.09 & 9.82 & 9.63 & 9.76 \\
$150\times160$ & 11.17 & 9.35 & 11.00 & 10.13 \\
$150\times170$ & 10.45 & 8.50 & 9.65 & 9.83 & \\
\hline
$170\times40$ & 9.13 & 6.23 & 6.08 & 5.30 \\
$170\times50$ & 8.11 & 7.40 & 6.58 & 6.15 \\
$170\times70$ & 7.48 & 7.24 & 6.90 & 7.13 \\
$170\times80$ & 8.03 & 7.76 & 7.45 & 7.71 \\
$170\times100$ & 9.08 & 8.31 & 8.47 & 8.06 \\
$170\times110$ & 9.18 & 8.44 & 8.53 & 8.35 \\
$170\times130$ & 9.62 & 8.51 & 8.45 & 8.67 \\
$170\times140$ & 9.83 & 8.29 & 8.59 & 8.91 \\
$170\times160$ & 9.87 & 8.34 & 9.55 & 9.15 \\
$170\times170$ & 10.09 & 8.50 & 9.74 & 9.56 \\
\hline
Total & 9.66 $\pm$ 2.29 & \textbf{8.85 $\pm$ 2.92} & 9.01 $\pm$ 2.92 &  \underline{8.96 $\pm$ 2.90} \\
\hline
\end{tabular}
\label{tab:mse}
\end{table}
\subsection{Scatter correction on the simulated reconstructions}
Table~\ref{tab:mse}  presents the RMSE (HU) for 30 different FOM sizes calculated for the reconstructed images. The lowest average RMSE was 8.85 ± 2.92, achieved by the net-$160\times128$, closely followed by the net-$40\times32$ with an RMSE of 8.96 ± 2.90. The net-$320\times256$ performed the worst, with an RMSE of 9.66 ± 2.29, followed by the net-$80\times64$ with an RMSE of 9.01 ± 2.92.

Exemplary axial images, corrected with different methods, along with the absolute difference from the corresponding MC-simulated ground truth are shown in Fig.~\ref{fig:sim_img}. Images with both the maximum $170\times170$ mm and minimum $130\times30$ mm FOMs from three different reconstructions are visualized. 
\begin{figure}[!htbp]
\centering
\includegraphics[width=\textwidth]{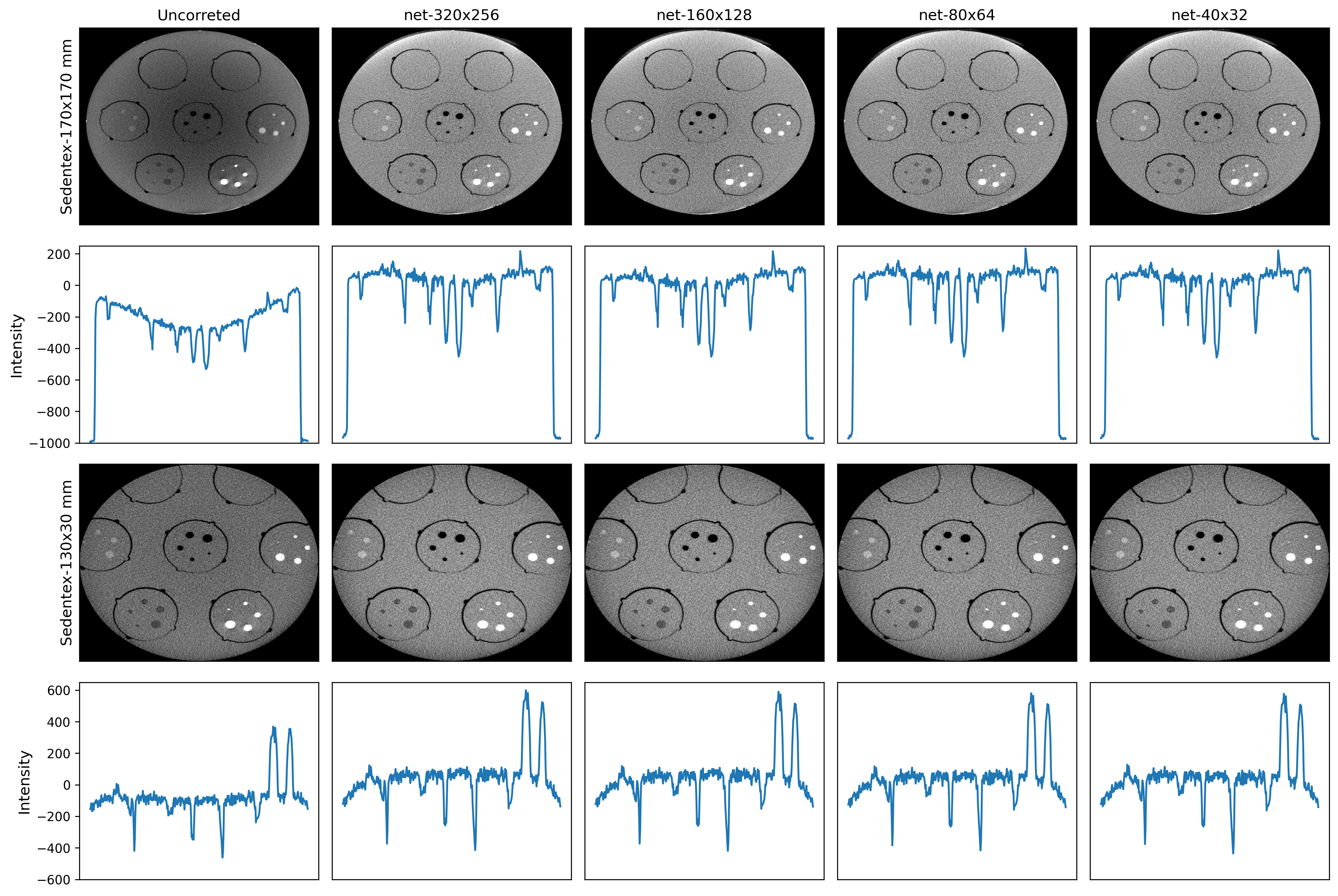}
\caption{SedentexCT IQ phantom scans. The first and third row show axial images from $170\times170$ and $130\times30$ mm FOMs, respectively. The display window center is 250 HU and width is 1500 HU. While the second and fourth row illustrate the mean HU intensity plots for a rectangular area placed at the center.}
\label{fig:img-sedentex}
\end{figure}
\subsection{Real scans}
\begin{figure}[!htbp]
\centering
\includegraphics[width=\textwidth]{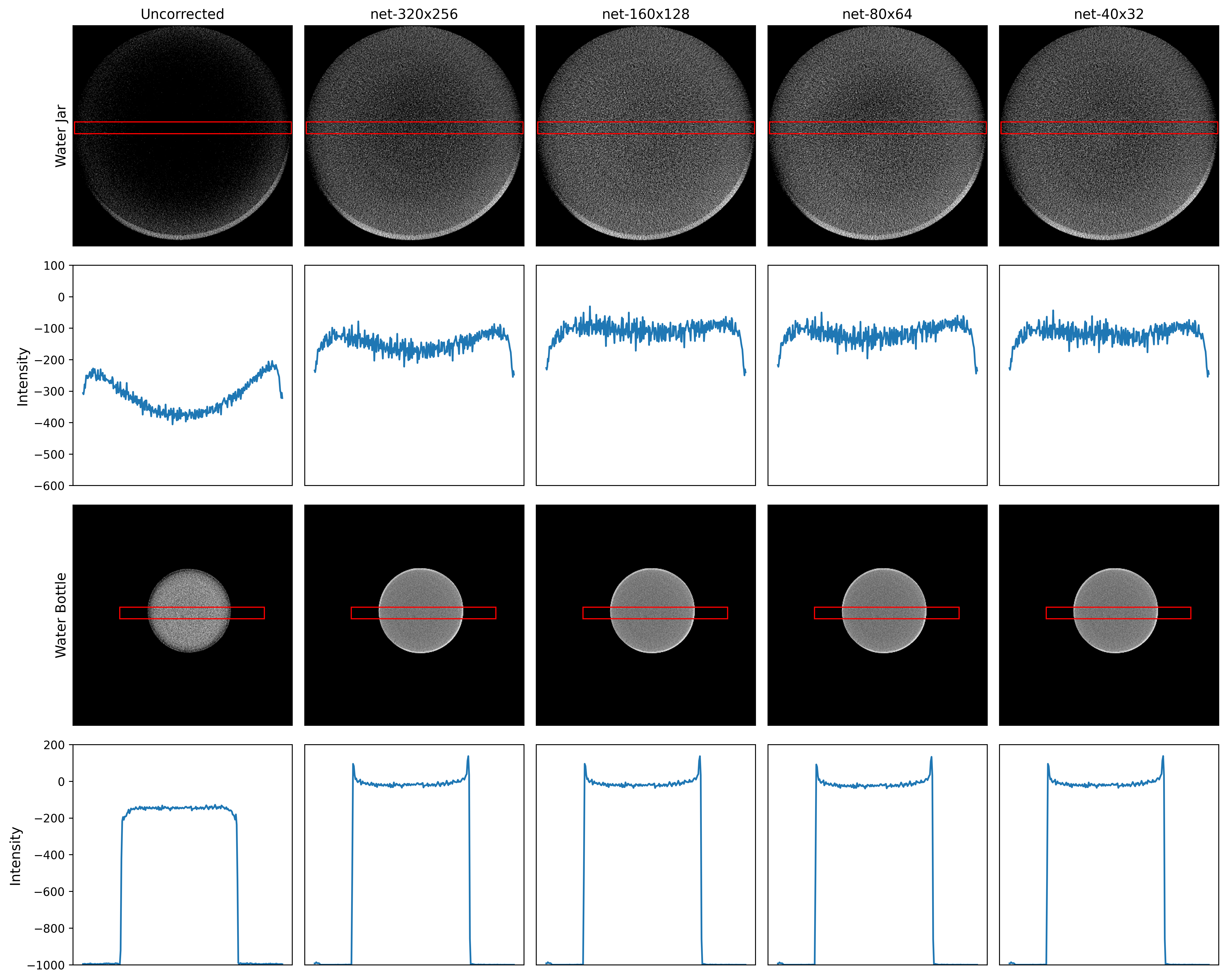}
\caption{Water phantom scans. First and third row images are from the scans of water jar and a small water bottle, respectively. The display window center is 0 HU and width is 500 HU. While the second and fourth row illustrate the mean HU intensity plots for window demonstrated by the colored rectangular area.}
\label{fig:img-water}
\end{figure}

Figure \ref{fig:img-water} shows the image reconstructions after scatter correction along with the uncorrected images for the scans of a large water jar and a small water bottle. For the large water jar, the effect of scatter artifact is considerably visible in the form of lower HU values and cupping artifact. All the networks improved the HU values with net-$160\times128$ and net-$40\times32$ showing least remaining cupping. On the other hand, the uncorrected image of water bottle shows no-cupping artifact and a smaller HU value inaccuracy. In this case, all the network corrected images have improved the HU value accuracy and improved the visibility of the edge of the phantom.

Figure~\ref{fig:img-sedentex} demonstrates two axial slices of uncorrected and scatter corrected reconstructions for the CBCT scans of SedentexCT IQ phantom scanned for $170\times170$ mm and $130\times30$ mm FOM sizes. For the scan of larger $170\times170$ mm FOM, the uncorrected axial slice has large shading and cupping artifact. All the neural network-based correction methods were reduced these artifacts. On the other hand, the uncorrected scan with $130\times30$ mm FOM, has smaller cupping and shading artifacts, however, HU values are still inaccurate. All the neural network-based methods improved the HU values and the contrast between different inserts.

\section{Discussion}\label{sec12}

We proposed to utilize the low-frequency nature of the scatter signal to downsample the scatter-corrupted images by a large factor. We empirically found that the bicubic interpolation performs best. This resulted in an ultrafast scatter estimation using a deep learning neural network. We also reduced the number of parameters in the network to account for the reduced size of the input, reducing the resource requirements further. We have demonstrated the effectiveness of the proposed method using a state-of-art scatter estimation network (Aux-Net \cite{agrawal2024}) on a large, simulated dataset as well as on real phantom scans.
\begin{figure}
    \centering
    \includegraphics[width=0.5\linewidth]{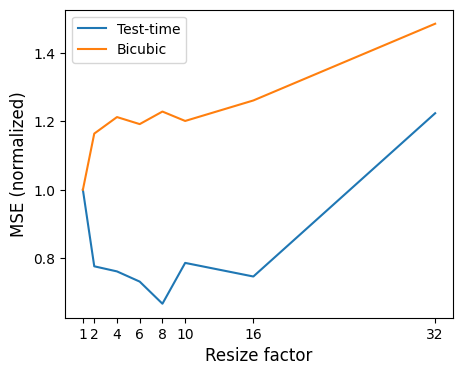}
    \caption{The test-time MSE errors vs. interpolation MSE errors on training data. }
    \label{fig:test-interp-error}
\end{figure}
In the first experiment, we compared the average scatter reconstruction error for the four interpolation methods by downsampling the scatter signal to different sizes and upsampling it back to the original size. We found bicubic interpolation to perform best, which is consistent with the finding in \cite{hirahara2021effect,hindratno2023impact}. For the bicubic interpolation, while the error started increasing immediately for a factor of two onward, the test error on the trained models, started increasing only after a factor of 8 (the size of $40\times32$), as shown in Fig.~\ref{fig:test-interp-error}. This finding indicates that an effective scatter estimation can be achieved at a much lower resolution size.
\begin{figure}[H]
    \centering
    \begin{minipage}{0.45\textwidth}
        \centering
        \begin{subfigure}[b]{\textwidth}
            \includegraphics[width=\textwidth]{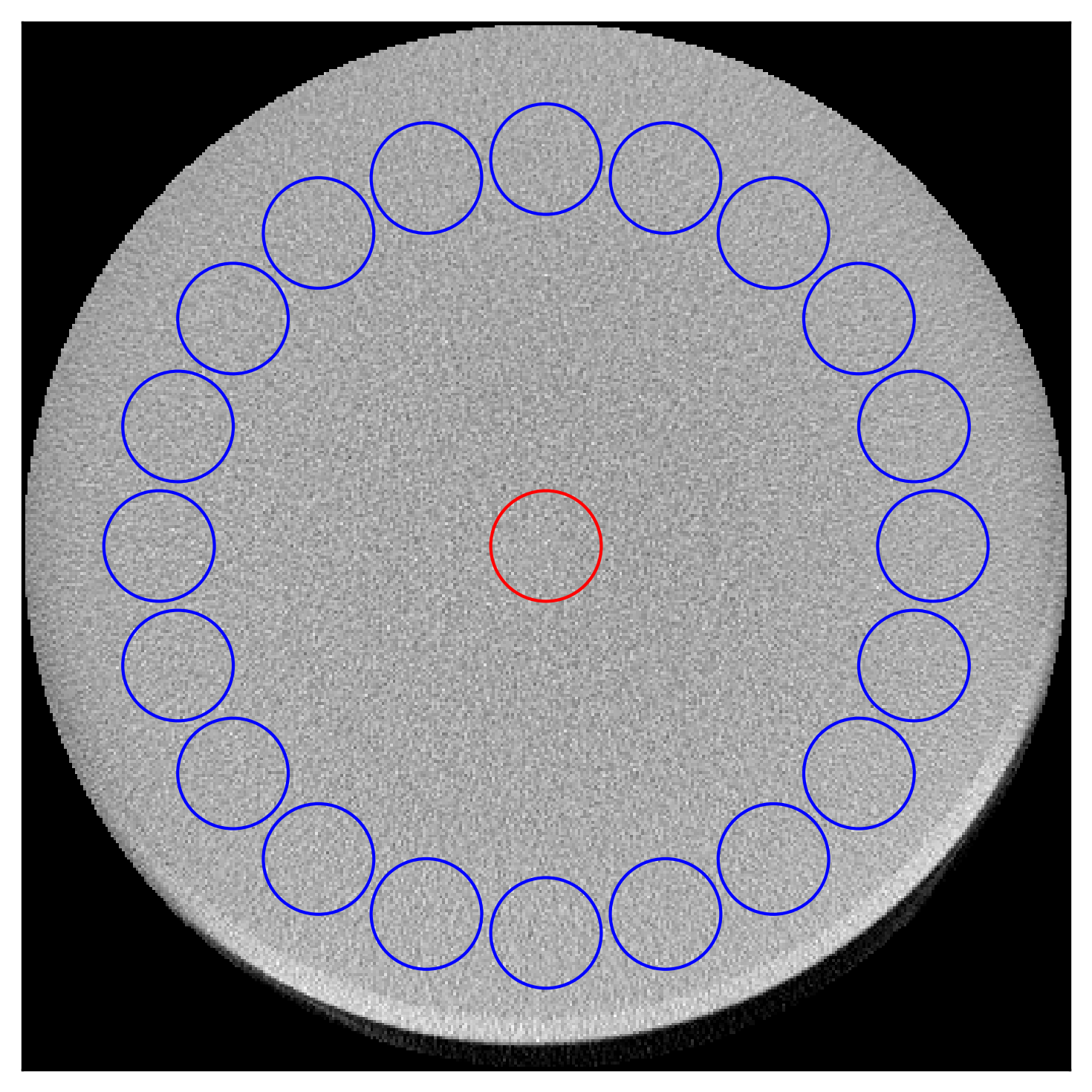}
            \caption{Water jar axial image showing the location of the central ROI (red) and 20 ROIs (blue) in the periphery, used to calculate image uniformity.}
        \end{subfigure}
        \begin{subfigure}[b]{\textwidth}
            \includegraphics[width=\textwidth]{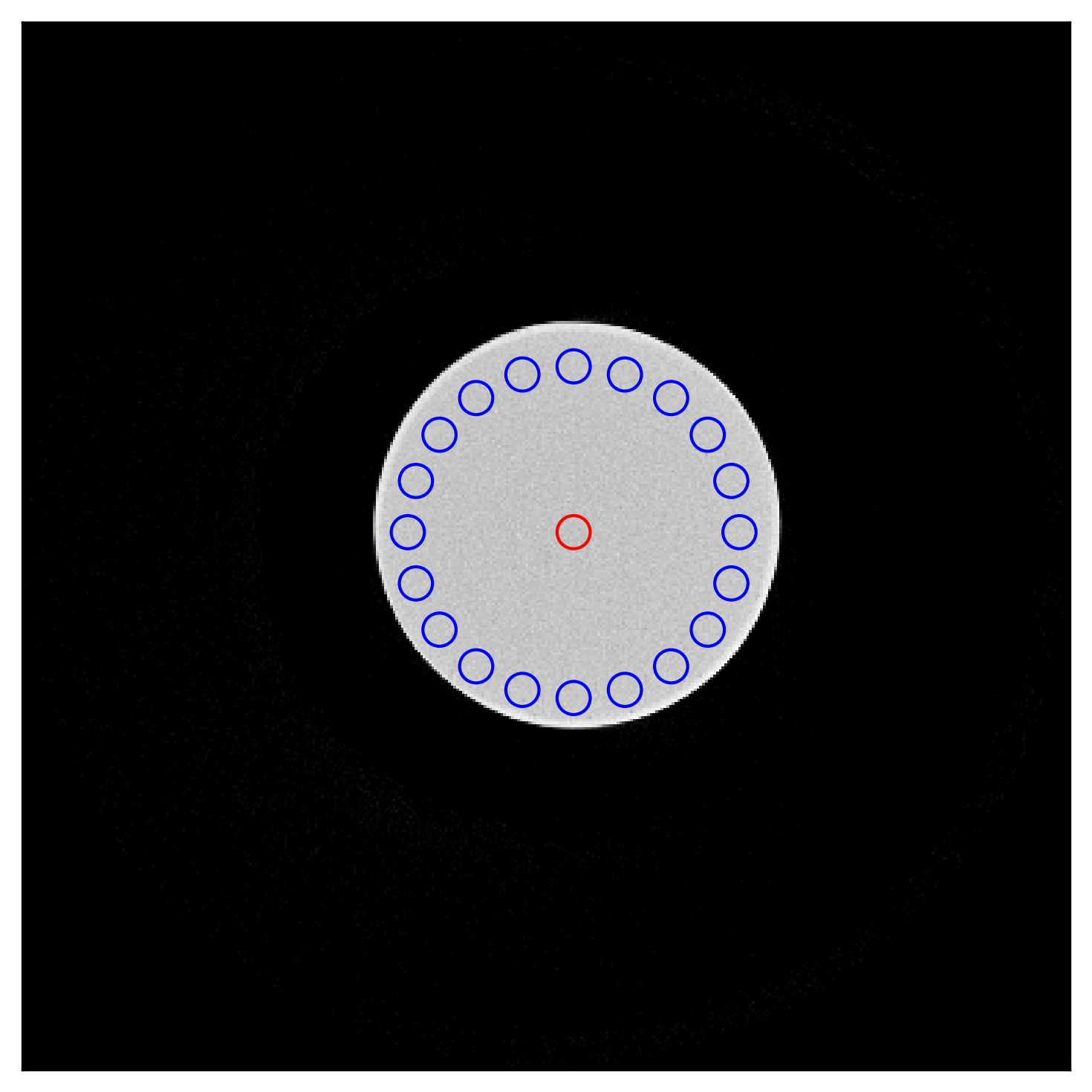}
            \caption{Middle axial slice from the water bottle scan showing the location of the central ROI (red) and 20 ROIs (blue) in the periphery, used to calculate image uniformity.}
        \end{subfigure}
    \end{minipage}
    \hfill
    \begin{minipage}{0.45\textwidth}
        \centering
        \begin{subtable}[b]{\textwidth}
            \centering
            \caption{Uniformity (HU) calculated for various methods for the selected ROIs from the water jar scan.}

            \begin{tabular}{ll}
                \toprule
                method & uniformity \\
                \midrule
                uncorrected & 118 \\
                net-$320\times256$ & 47 \\
                net-$160\times128$ & 18 \\
                net-$80\times64$ & 29 \\
                net-$40\times32$ & 18 \\
                net-$20\times16$ & 27 \\
                \bottomrule
            \end{tabular}
            
        \end{subtable}
        \begin{subtable}[b]{\textwidth}
            \centering
            \caption{Uniformity (HU) calculated for various methods for the selected ROIs from the water bottle scan.}
            \begin{tabular}{ll}
                \toprule
                 method & uniformity \\
                \midrule
                uncorrected & 8 \\
                net-$320\times256$ & 14 \\
                net-$160\times128$ & 17 \\
                net-$80\times64$ & 27 \\
                net-$40\times32$ & 16 \\
                net-$20\times16$ & 18 \\
                \bottomrule
            \end{tabular}
        \end{subtable}
    \end{minipage}
    \caption{Image uniformity in the corrected images compared for the different methods.}
    \label{fig:uniformity}
\end{figure}
Further, we compared the scatter estimation performance of five different networks, trained with five input sizes for the simulated projections. The net-$40\times32$ outperformed all the other networks showing least MSE ($0.134 \pm 0.009$) and comparable MAPE ($3.85 \pm 0.10$) with only 310 MB of GPU memory, processing a batch size of 64 in just under 6 ms. In comparison, net-$320\times256$ took 90 ms with 3.9 GB GPU memory and had worse MAPE ($4.42 \pm 0.18$ and MSE ($0.20 \pm 0.014$). Similar results were obtained for the simulated reconstructions after the scatter correction. Specifically, net-$40\times32$ had second lowest RMSE ($8.96 \pm 2.90$), which was better than the RMSE ($9.66 \pm 2.29$) of net-$320\times256$. However, net-$160\times128$ had slightly better RMSE of $8.85 \pm 2.92$. It is clear that the smaller input size significantly reduces the computational requirements while maintaining acceptable error rates. Which shows that the net-$40 \times 32$ configuration offers a good balance between performance and resource efficiency.
Additionally, the uniformity is shown for the water jar and water bottle scans in Fig. \ref{fig:uniformity}. The difference between the uniformity values of the uncorrected images is significant, with the uniformity being 118 HU for the large water jar and 8 HU for the small water bottle. This discrepancy results from the large amount of scatter in the larger water jar compared to the smaller water bottle. An interesting observation is that while the neural network-based methods improved the uniformity in the water jar scan (18 HU), the uniformity worsened for the water bottle (16 HU). This can be attributed to the over-correction in the water bottle compared to the water jar since the training data did not include such small objects. Nonetheless, the net-$40\times32$ achieved comparable results to the other best performing networks for both scans.
\section{Conclusion}\label{sec13}
Leveraging the low-frequency nature of scatter distribution to train at a much lower resolution accelerated the inference for deep learning-based scatter estimation. The performance of the method is comparable with the state-of-art scatter estimation network, trained with higher resolution inputs. These findings enhance the potential of the deep learning-based scatter estimation in low-resource environments like mobile CBCT or edge devices without compromising the image quality. In resource-rich environments, it remains beneficial to reduce the total time required to generate scatter-corrected projections. Our future research will involve applying the downsampling in the angular dimension as well, and investigate the applicability of a three-dimensional network, giving the network more context and making the inference faster.

\backmatter

\section*{Declarations}
\begin{itemize}
\item \textbf{Funding} The work is in-part supported by a Business Finland grant (number: 8175/31/2022) under the TomoHead project.
\item \textbf{Conflict of interest} Harshit Agrawal and Ari Hietanen are employees at Planmeca Oy.
\item \textbf{Author contribution} All authors contributed to the study. HA: Conceptualization, Experimentation, and Writing – original draft. AH: Writing – review \& editing. SS: Conceptualization, Writing – review \& editing.
\end{itemize}

\begin{appendices}
\section{Appendix}\label{secA1}
\begin{table}[h]
\caption{The 18 FOM sizes used for the simulated training dataset are shown. The FOM sizes are gievn in terms of  diameter (mm) $\times$ height (mm).}

\centering   
\begin{tabular}{|c|*{6}{c|}}
\hline
 120$\times$30 & 120$\times$60 & 120$\times$90 & 120$\times$120 & 120$\times$150 & 120$\times$180 \\
\hline
140$\times$30 & 140$\times$60 & 140$\times$90 & 140$\times$120 & 140$\times$150 & 140$\times$180 \\
\hline
160$\times$30 & 160$\times$60 & 160$\times$90 & 160$\times$120 & 160$\times$150 & 160$\times$180 \\
\hline
\end{tabular}

\end{table}

 \begin{table}[ht]
\caption{The 30 FOM sizes used for the simulated testing dataset are shown in terms of diameter (mm) $\times$ height (mm).}

\centering   
\begin{tabular}{|c|*{10}{c|}}
\hline
 130$\times$40 & 130$\times$50 & 130$\times$70 & 130$\times$80 & 130$\times$100 & 130$\times$110 & 130$\times$130 & 130$\times$140 & 130$\times$160 & 130$\times$170\\
\hline
 150$\times$40 & 150$\times$50 & 150$\times$70 & 150$\times$80 & 150$\times$100 & 150$\times$110 & 150$\times$130 & 150$\times$140 & 150$\times$160 & 150$\times$170\\
\hline
 170$\times$40 & 170$\times$50 & 170$\times$70 & 170$\times$80 & 170$\times$100 & 170$\times$110 & 170$\times$130 & 170$\times$140 & 170$\times$160 & 170$\times$170 \\
\hline
\end{tabular}

\end{table}

\end{appendices}

\bibliographystyle{unsrt}
\bibliography{main}
\end{document}